# Gate-tuned quantum Hall states in Dirac semimetal (Cd$_{1-x}$Zn$_x$)$_3$As$_2$

**Authors**

Shinichi Nishihaya[1], Masaki Uchida[1]*, Yusuke Nakazawa[1], Markus Kriener[2], Yusuke Kozuka[1], Yasujiro Taguchi[2], Masashi Kawasaki[1,2]

**Affiliations**

[1] Department of Applied Physics and Quantum-Phase Electronics Center (QPEC), the University of Tokyo, Tokyo 113-8656, Japan

[2] RIKEN Center for Emergent Matter Science (CEMS), Wako 351-0198, Japan

*To whom correspondence should be addressed: E-mail: uchida@ap.t.u-tokyo.ac.jp

**One-sentence summary**

Electronic structure of quantum confined Dirac semimetal is elucidated by establishing carrier control techniques in films.

**Abstract**

The recent discovery of topological Dirac semimetals (DSM) has provoked intense curiosity not only on Weyl physics in solids, but also on topological phase transitions originating from DSM. One example is controlling the dimensionality to realize two-dimensional quantum phases such as quantum Hall and quantum spin Hall states. For investigating these phases, the Fermi level is a key controlling parameter. From this perspective, we report here the carrier-density control of quantum Hall states realized in thin films of DSM Cd$_3$As$_2$. Chemical doping of Zn combined with electrostatic gating has enabled us to tune the carrier density over a wide range and continuously even across the charge neutrality point. Comprehensive analyses of the gate-tuned quantum transport have revealed Landau level formation from linearly dispersed sub-bands and its contribution to the quantum Hall states. Our achievements pave the way also for investigating the low energy physics near the Dirac points of DSM.

**MAIN TEXT**

**Introduction**

In the past decade, electron transport in Dirac systems has attracted intensive research efforts to uncover various kinds of unprecedented phenomena in two-dimensional (2D) Dirac systems such as graphene (*1*) and the surface states of topological insulators (*2-4*). Recently, this playground has been extended even to three-dimensional (3D) systems after the experimental discovery of topological Dirac semimetals (DSM) (*5-7*) and Weyl semimetals (WSM) (*8-10*) in solids, where the bulk conduction band (CB) and valence band (VB) touch at finite pairs of Dirac or Weyl points to form 3D Dirac dispersions. Among the DSM materials, $Cd_3As_2$ has been treated as an ideal testbed, because of its ultra-high mobility and chemical stability in air, bearing potential for finding new topological phases. Soon after its theoretical prediction (*5*), the characteristic linear dispersion and the band inverted nature of DSM have been revealed by spectroscopy experiments (*6,11,12*). They have been subsequently followed by a surge of reports on transport phenomena including quantum oscillations (*13-15*), chiral anomaly (*16-18*), and Fermi arc mediated surface transport (*19*).

While the above experimental observations on the transport of DSM are mainly based on the 3D nature of bulk samples, a two-dimensionally confined Fermi surface achievable in a thin film sample also offers opportunities for realizing novel 2D quantum states. The recent observation of the quantum Hall effect (QHE) in $Cd_3As_2$ is one example (*20-22*). It has been revealed that due to the large sub-band splitting originating from the linear dispersion of $Cd_3As_2$, a 2D nature of the Fermi surface can easily emerge even with a moderate confinement (*20*). This aspect of $Cd_3As_2$ is highly beneficial for realizing another type of 2D quantum states called quantum spin Hall insulator as well, which is predicted to be achievable within the band inverted regime of $Cd_3As_2$ (*5,20,23*).

From the view point of transport measurements, however, the major obstacle in the pursuit of quantum phases in $Cd_3As_2$, regardless whether using 3D bulk or 2D confined films, is the defect-induced high carrier density, which shifts the Fermi level away from the charge neutral Dirac points. As $Cd_3As_2$ is easily electron doped due to As deficiencies incorporated during the synthesis processes, achieving a low electron density still remains challenging even in samples with high crystallinity (*13-15,19,20*). Since the band properties accessible through transport measurements are limited to those at the Fermi level, systematic carrier control of $Cd_3As_2$ is of crucial importance to elucidate its electronic structure in detail. In this regard, chemical doping of Zn is known to be an effective way to reduce the electron density of $Cd_3As_2$. By making a solid solution with $Zn_3As_2$, a *p*-type trivial semiconductor with a band gap of 0.93 ~ 1.1 eV (*24*), the residual electrons of $Cd_3As_2$ can be effectively compensated (*25,26*). The only but serious drawback of the use of Zn doping, though, is that it alters the band structure as heavy doping would lead to a suppression of the band inversion in $Cd_3As_2$ and hence loss of its DSM nature (*26*). In the sense of avoiding such heavy doping effect, electrostatic carrier depletion in a field effect transistor configuration is also a useful choice in the case of thin film samples. So far, gate-tuning using an electric-double-layer transistor has been reported on rather thick $Cd_3As_2$ films (~ 50 nm) (*27*). Capability of tuning the Fermi level position would offer great advantages in studying the 2D quantum transport and clarifying the actual band structure of confined films as well.

In this context, we report quantum transport of two-dimensionally confined $Cd_3As_2$ films where the carrier density is controlled over a wide range using the combination of Zn doping and electrostatic gating (Fig.1 (A)). The effective carrier suppression with minimal Zn doping makes it possible to keep the high mobility and the linear dispersion of $Cd_3As_2$. The 2D nature of the confined Fermi surface emerges as the electron density is reduced, leading to the observation of clear quantum Hall (QH) states with various filling factors. The detailed analysis of the QH states while tuning the Fermi level position have revealed Landau level formation from linearly dispersed sub-bands and their occupation transitions depending on the Fermi level position, which in turn provides a clearer understanding of the band structure and the confinement effect in $Cd_3As_2$ films.

**Results**

**Zn doping and field effect in $Cd_3As_2$ thin films**

Zn doped $Cd_3As_2$ thin films were fabricated on $SrTiO_3$ (100) substrate by a combination of pulsed laser deposition and thermal treatment, following the same procedures explained in ref. 20 (see also Materials and Methods). From x-ray diffraction measurements, the $(Cd_{1-x}Zn_x)_3As_2$ alloy films are confirmed to have a (112) oriented single phase and their lattice constant agrees well with Vegard's law (25) (Fig. S1). The films were deposited through a stencil metal mask into a Hall bar shape with a typical channel width of 60 μm, and the $SrTiO_3$ substrates were directly used as gate dielectric in a back-gate configuration as depicted in Fig. 1(B).

As Zn is doped in $Cd_3As_2$, the temperature dependence of the longitudinal resistance $R_{xx}$ exhibits a transition from metallic to semiconducting behavior (Fig. 1(C)). The electron density of the $(Cd_{1-x}Zn_x)_3As_2$ films is shown as a function of the Zn concentration $x$ in Fig. 1(D) along with the electron density reported for bulk single crystals for comparison (26). A clear suppression of the residual electron density is observed as the Zn concentration increases. It is notable that the electron density of thin films is suppressed by Zn doping far more effectively than that of bulk single crystals, reflecting differences in the growth methods (see also the section S1 in the Supplementary Materials). The electron mobility shown in Fig. 1(E) keeps high values exceeding $10^4$ $cm^2$/Vs for thin films, which is similar to those of bulk samples (26). This high mobility of the films in the low electron density region makes it possible to investigate quantum transport phenomena down to the quantum limit.

Effective carrier suppression with less doping amount *x* is preferable in terms of keeping the Dirac dispersion of $Cd_3As_2$. Since $Zn_3As_2$ is a semiconductor with ordinary band ordering, doping Zn into $Cd_3As_2$ not only compensates electrons but also lifts up the band inversion, finally inducing a topological phase transition from a DSM to a trivial semiconductor. The critical composition $x_c$ for this transition is predicted to be around 0.17 based on the early magneto-optical measurements and theoretical studies (*28*) reporting that the energy gap of $(Cd_{1-x}Zn_x)_3As_2$ varies linearly with *x* and the band inversion energy for $Cd_3As_2$ is 0.19 eV. Therefore, it is reasonable to conclude that at least the *x* = 0.06 and 0.11 samples are still in the band inverted regime as depicted in the schematics of Fig. 1(D), and our magnetotransport data of the *x* = 0.11 sample presented later indeed reflect the existence of a linear energy dispersion.

Figure 2(A) presents the magnetic-field dependence of longitudinal resistance $R_{xx}$ and Hall resistance $R_{yx}$. With the progressive suppression of the electron density, the amplitudes of the quantum oscillations are enhanced in $R_{xx}$ and QH states start to appear with characteristic quantized plateaus in $R_{yx}$. The observation of the QHE, which is essentially a 2D phenomenon, even in 35 nm thick films can be explained by the large Fermi velocity in the confinement direction. The sub-band splitting energy becomes as large as 80 meV in the case of the 35 nm thick film, for instance, in the *x* = 0.11-doped sample as estimated later. In such a situation, only a few sub-bands are occupied by electrons at low carrier densities, and thus the 2D nature of each sub-band emerges.

The observed Hall plateaus in $(Cd_{1-x}Zn_x)_3As_2$ films exhibit quantization at $\sigma_{xy} = sn\frac{e^2}{h} = \nu\frac{e^2}{h}$, where *n* is integers, *s* is the spin and valley degeneracy, and $\nu$ is the filling factor, rather than at the half integer shifted values $\sigma_{xy} = s\left(n + \frac{1}{2}\right)\frac{e^2}{h}$ typically observed in gapless Dirac systems (*29,30*). Lack of such a shift in $\sigma_{xy}$ indicates the absence of the zero energy *N* = 0 Landau level, which is naturally expected from the existence of a confinement gap in the sub-band (*20*). The energy dispersion on a momentum plane off the Dirac points is well described by a hyperbolic function with a finite gap at the bottom (*5*) (Fig. 1(A)), which has been resolved in the angle-resolved photoemission spectroscopy experiment of bulk $Cd_3As_2$ (*6*). In this sense, despite the confinement gap, the linear character of the energy dispersion is still maintained in the current system, and intriguingly the observed QH states are well explained by Landau quantization of such a gapped Dirac dispersion.

Combined with field effect, the $(Cd_{1-x}Zn_x)_3As_2$ films can be depleted further and even gate-tuned into a *p*-type regime, as indicated by a clear sign change in $R_{yx}$ of the negatively biased $x = 0.17$ sample. Figure 2(B) presents the gate voltage $V_G$ dependence of $R_{xx}$ and the sheet carrier density. Following the carrier type inversion, a peak structure appears in $R_{xx}$, which corresponds to the charge neutrality point (CNP). The resistance value at the CNP exhibits a diverging behavior with *x*. The suppressed band inversion of $Cd_3As_2$ by Zn doping and hence the increase of the energy gap in the confined sub-band accounts for such a dependence. In the *p*-type regime for $x = 0.11$ and 0.14, $R_{yx}$ exhibits multi-carrier conduction in contrast to single-hole conduction for $x = 0.17$ (Fig. S2). The existence of residual electrons implies either the activation of electrons or the formation of electron-hole puddles (*31*). The deviation from a linear $V_G$ dependence of the electron density around the CNP may also be related to such an inhomogeneity effect. On the other hand, Zn-doping is expected to enhance the energy gap and hinder the puddle formation, eventually leading to single-hole conduction for $x = 0.17$. Both the quantum oscillations and the QH states are suppressed in the *p*-type region probably due to the lower hole mobility or the heavier mass of the valence band (Fig. S2), which is consistent with previous reports (*27,32*).

**Gate-modulation of quantum Hall states**

Having established the carrier-density control by field effect, we now investigate quantum transport at different Fermi level positions. In Fig. 3, we summarize the results for $x = 0.11$, where the mobility is the highest among the samples. As $V_G$ is negatively increased in −5 V steps, Hall

*Science Advances*       Manuscript Template        Page **6** of **20**

plateaus with smaller filling factors $\nu$ show up one by one, and finally the system reaches the quantum limit at $V_G = -20$ V, where all the electrons fall into the lowest Landau level for magnetic fields larger than $B = 8$ T. The magnetic field dependence of the QH states reveals that $\nu$ changes by 2 or 4 depending on the field, except for the last transition from $\nu = 2$ to 1. With the Fermi level located above the saddle point, the system only possesses a spin degeneracy of $s = 2$. Therefore, discrete jumps by 4 in $\nu$, for example around $B = 7$ T at $V_G = -5$ V, strongly suggest the existence of other sub-bands contributing to the QH states. To obtain a detailed picture of the Landau level occupation and the resulting QH states when multiple sub-bands are involved, information about the band dispersion in $(Cd_{1-x}Zn_x)_3As_2$ films needs to be extracted.

For this purpose, the oscillation terms of $R_{xx}$ in lower fields are analyzed to estimate band properties such as Fermi momentum $k_F$, effective mass $m^*$ and Fermi velocity $v_F$. The Fermi momentum $k_F$ and the Fermi surface area $S_F = \pi k_F^2$ are extracted from the periodicity of quantum oscillations using Fourier transformation. We note that the oscillations are mainly dominated by the single periodicity originating from the main sub-band (referred to as SB1), and the second occupied sub-band (referred to as SB2) is hardly resolved in the Fourier spectrum (Figs. S3 and S4). The effective mass $m^*$, on the other hand, can be estimated from the damping behavior of the oscillation amplitudes with temperature. Since the dielectric constant of the $SrTiO_3$ substrates highly depends on temperature (33), the electron density was kept constant by tuning $V_G$ at each temperature to investigate the temperature dependence of the oscillations (Fig. S3). Figure 3(C) shows a typical example of the oscillatory components of $R_{xx}$ at different temperatures. Here, $V_G^0$ denotes the bias voltage at 2 K, and at higher temperatures $V_G$ was adjusted to keep the electron density of the sample constant as deduced by low-field Hall measurements.

Figure 3(D) presents the change of $m^*$ as a function of $k_F$ for the main sub-band SB1. The effective mass exhibits a decreasing trend as the Fermi surface area shrinks. This behavior contrasts with the case of a parabolic band where $m^*$ is independent of the Fermi surface area, suggesting that the energy dispersion remains linear for the $x = 0.11$ sample as expected. The variation of the extracted effective mass at each Fermi level may be ascribed to the influence of the second occupied sub-band SB2, which may affect the appearance of oscillation amplitudes and bias the estimated $m^*$ in either direction, underestimation or overestimation (see also section S4 in the Supplementary Materials). By taking the average of these $m^*$, the Fermi velocity $v_F$ of SB1 is estimated to be about $1.1 \times 10^6$ m/s, which is close to typical values previously reported for bulk $Cd_3As_2$ samples (*13-15*).

**Analysis of Landau level occupations**

Having $k_F$ of SB1 determined, the carrier occupation of SB2 can be quantified as the difference between total electron density and partial electron density $n_{SB1} = k_F^2/2\pi$ of SB1. Figure 4(A) displays the electron density obtained from different experimental data as a function of $V_G$; $n_{Hall}$ from the Hall coefficient shown in Fig. 2(B), $n_{QHE}$ from the quantization periodicity of the QH states shown in Fig. 4(B), and $n_{SB1}$ from $k_F$ of SB1. Both $n_{Hall}$ and $n_{QHE}$ correspond to the total electron density of the sample and thus coincide over the entire range of $V_G$. On the other hand, $n_{SB1}$ shows smaller values than the other two at $V_G > -15$ V, clearly indicating the occupation of SB2. The occupation of the third sub-band (referred to as SB3) is also possible, but electrons in SB3 should be rapidly depleted under high magnetic fields. Therefore, for $V_G > -15$ V, it is reasonable to assume that two sets of Landau levels originating from SB1 and SB2 are involved in the emergence of the QH states.

Landau level occupation within such multiple sub-bands can be well understood by plotting the filling factor $\nu$ against the inverse of the magnetic field as presented in Fig. 4(B). Because each Landau level possesses a density of states of $\frac{eB}{h}$ at a given magnetic field $B$ and total electron density $n$, quantized plateaus appear according to the relationship $n = \nu \frac{eB}{h}$. Thus, a linear fit to the data points yields the total electron density $n_{QHE}$ as plotted by solid lines in Fig.

4(A), regardless of the details of the multiple sub-bands (*34*). Next, we assume only the presence of SB1 and draw straight lines based on the Fermi surface area of SB1 from each data point, shown as broken lines in Fig. 4(B) (see also Fig. S6). In contrast to the previous case, they do not pass through the origin but have a finite intercept on the horizontal axis. This is because the filled Landau levels and the corresponding density of states in SB2 are ignored. From this viewpoint, it is reasonable to consider that the remaining intercept on the horizontal axis in return provides a rough estimation of the number of ignored Landau levels in SB2. For instance, the intercept of the broken line in the $V_G = -5$ V case is 2 for the $\nu = 6$ state, and this can be interpreted as the occupation of one Landau level from SB2 in addition to two Landau levels from SB1 to account for the $\nu = 6$ state. In this way, the detailed breakdown of the filling factors between SB1 and SB2 at each magnetic field can be obtained (Fig. S6).

Figure 4(C) summarizes the magnetic field dependence of the Landau level occupation at each gate voltage. The Landau level energies of confined $Cd_3As_2$ with a hyperbolic dispersion can be formulated as follows by applying the semi-classical quantization condition to the hyperbolic dispersion (see also the section S6 in the Supplementary Materials),

$$E_N = \hbar v_F \sqrt{\frac{2e(N+\gamma)B}{\hbar} + \left(\frac{E_G}{\hbar v_F}\right)^2}.$$

Here, $N$ is the Landau index and $v_F$ is the Fermi velocity in the in-plane direction ($\perp$[112]). $\gamma$ reflects the Berry phase $\Phi$ and is defined as $\gamma = \frac{1}{2} - \frac{\Phi}{2\pi}$. We simply assume $\gamma = 0.5$ for the trivial case because of the gapped dispersion. $E_G$ is the energy difference between the bulk Dirac points and the bottom of each sub-band, corresponding to the confinement-induced gap. Based on the information of electron density, number of occupied Landau levels in SB1, SB2, and field positions of the plateau transitions, the following parameters are estimated to agree well with the experimental observations; $E_{G,SB1} = 60$ meV, $E_{G, SB2} = 140$ meV, $v_{F,SB1} = 1.1 \times 10^6$ m/s as determined in Fig. 3(D), $v_{F,SB2} = 8\times10^5$ m/s (see section S6 in the Supplementary Materials for a more detailed description of the estimation procedure). The occupation of SB2 above $E = 140$ meV opens the possibility that the Landau levels overlap at the Fermi level and hence certain filling factors are skipped. To account for the plateau transition from $\nu = 2$ to $\nu = 1$, spin-split levels are calculated by introducing the Zeeman term $\mu_B g B$ with $g = 25$ (*12*). Moreover, based on Fig. 4(C), the sub-band splitting energy ($E_{G,SB2} - E_{G,SB1}$) reaches 80 meV. Taking into account the confinement thickness of 35 nm, the Fermi velocity in the confinement direction ($v_{F//[112]}$) can be estimated to be around $7\times10^5$ m/s. Thus, combined with the systematic Fermi level control, the band structure of the original 3D bulk is also accessible through the comprehensive analysis of QH states in 2D confined films.

**Discussion**

Here, we assume the sub-band splitting energy is kept constant over the investigated range of gate-voltage and magnetic field for quantitative analysis, and it well captures the essential picture of the experimentally observed QH states. For more strict quantification, a self-consistent calculation is preferable because the possible charge transfer between the spatially separated wave functions of the sub-bands can also induce a shift of sub-band splitting energy (*35*). Compared to the GaAs/Al$_x$Ga$_{1-x}$As wide well case (*35*), however, we note that the possible shift (~ a few meV) only has negligible effects because of the large confinement-induced sub-band splitting (80 meV) in the Dirac dispersion of (Cd$_{1-x}$Zn$_x$)$_3$As$_2$.

In addition to the detailed analysis of the QH states, systematic chemical doping and carrier control in thin films of Cd$_3$As$_2$ provide new opportunities to investigate the exotic phenomena predicted in the DSM. As the Zn-concentration dependence of the peak resistance at the CNP in Fig. 2B clearly reveals the suppression of band inversion in Cd$_3$As$_2$, Zn doping is useful in modulating the band structure of DSM. In particular, the controllability of the Dirac point positions in the momentum space by Zn doping provides an additional controlling parameter for investigating the topological transport phenomena such as chiral anomaly, spin Hall effect (*36*), Fermi arc mediated quantized transport (*37*), and potential topological superconductivity (*38,39*). The electrostatic carrier control is also useful for their investigations, as these phenomena appear dominantly when the Fermi level is close to the Dirac points.

While Zn doping induces the topological phase transition from 3D DSM to a trivial insulator, the phase transitions into other 2D quantum phases such as quantum spin Hall insulator and quantum anomalous Hall insulator is possible in quantum wells and heterostructures (*5,20,40*). For the detection of such quantum phases, tuning the Fermi level position into the bulk gap to single out the edge transport is inevitable. In this sense, our achievements on the quantitative analysis of the confinement effect and the carrier control across the CNP also lay the foundation for such attempts.

In summary, we have studied QH states of two dimensionally confined $Cd_3As_2$ films where the carrier density is systematically controlled by chemical doping with Zn and electrostatic gating. Benefiting from the comprehensive analysis down to the quantum limit, Landau level formation from the linearly dispersed multiple sub-bands and their occupation transitions depending on the Fermi level position have been revealed. Our work on the chemical doping and Fermi level tuning for $Cd_3As_2$ thin films also provide the important basis for the pursuit of transport manifestation of 3D Dirac Fermions and topological phase transitions predicted to occur in DSM.

**Materials and Methods**

**Film growth**

For the growth of $(Cd_{1-x}Zn_x)_3As_2$ thin films, $Cd_3As_2$ and $Zn_3As_2$ layers with a designed thickness ratio were first deposited in a stacking manner on $SrTiO_3$ substrates, which were then capped with $TiO_2$ (1 nm) / MgO (5 nm) / $Si_3N_4$ (200 nm). A patterned stencil metal mask was set on the substrate during the deposition process to make the film into the Hall bar shape. The $TiO_2$ and MgO capping layers were also deposited successively through the mask, while it was retracted when depositing the $Si_3N_4$ layer so that the latter covers the entire film including the Hall bar edges. All the layers were deposited using a pulsed laser deposition technique at room temperature. The films were then subject to annealing at 600 °C in air. This annealing process promotes crystallization of the films as well as the formation of a solid solution of the two arsenide layers. In the presence of the capping materials, temperature can be elevated without evaporation of the deposited arsenide films nor any undesired chemical reactions taking place, resulting in $(Cd_{1-x}Zn_x)_3As_2$ alloy films of high crystalline quality. The Zn concentration is simply controlled by modifying the thickness ratio of the deposited $Cd_3As_2$ and $Zn_3As_2$ layers.

**Transport measurements**

Low-temperature transport measurements were carried out in a Physical Property Measurement System (PPMS, Quantum Design) with a magnetic field up to 9 T and at a base temperature of 2 K. Hall measurements were conducted using a lock-in technique. The excitation current was kept constant at 0.1 µA with a frequency of 13 Hz. For field effect measurements, DC bias voltage was applied to the Cu gate electrode.

**Acknowledgments**

**General**: We acknowledge fruitful discussions with N. Nagaosa, Y. Tokura, R. Arita, Y. Yamaji, M. Hirschberger, and T. Liang. **Funding:** This work was supported by JST CREST Grant No. JPMJCR16F1, Japan, and by Grant-in-Aids for Scientific Research (C) No. JP15K05140 and Scientific Research on Innovative Areas "Topological Materials Science" No. JP16H00980 from MEXT, Japan. **Author contributions:** S.N., M.U., and M.Kawasaki designed the experiments. S.N., M.U., Y.N. synthesized the bulk targets with M.Kriener and performed thin film growth, device fabrication and low temperature transport measurements. S.N. analyzed the data and wrote the manuscript with contributions from all the authors. Y.K., Y.T., and M.Kawasaki jointly discussed the results. M.U. and M.Kawasaki conceived the project. All authors have approved the final version of the manuscript. **Competing interests:** The authors declare that they have no competing interests. **Data and materials availability:** All data needed to evaluate the conclusions in the paper are present in the paper and/or the Supplementary Materials. Additional data related to this paper may be requested from the authors.

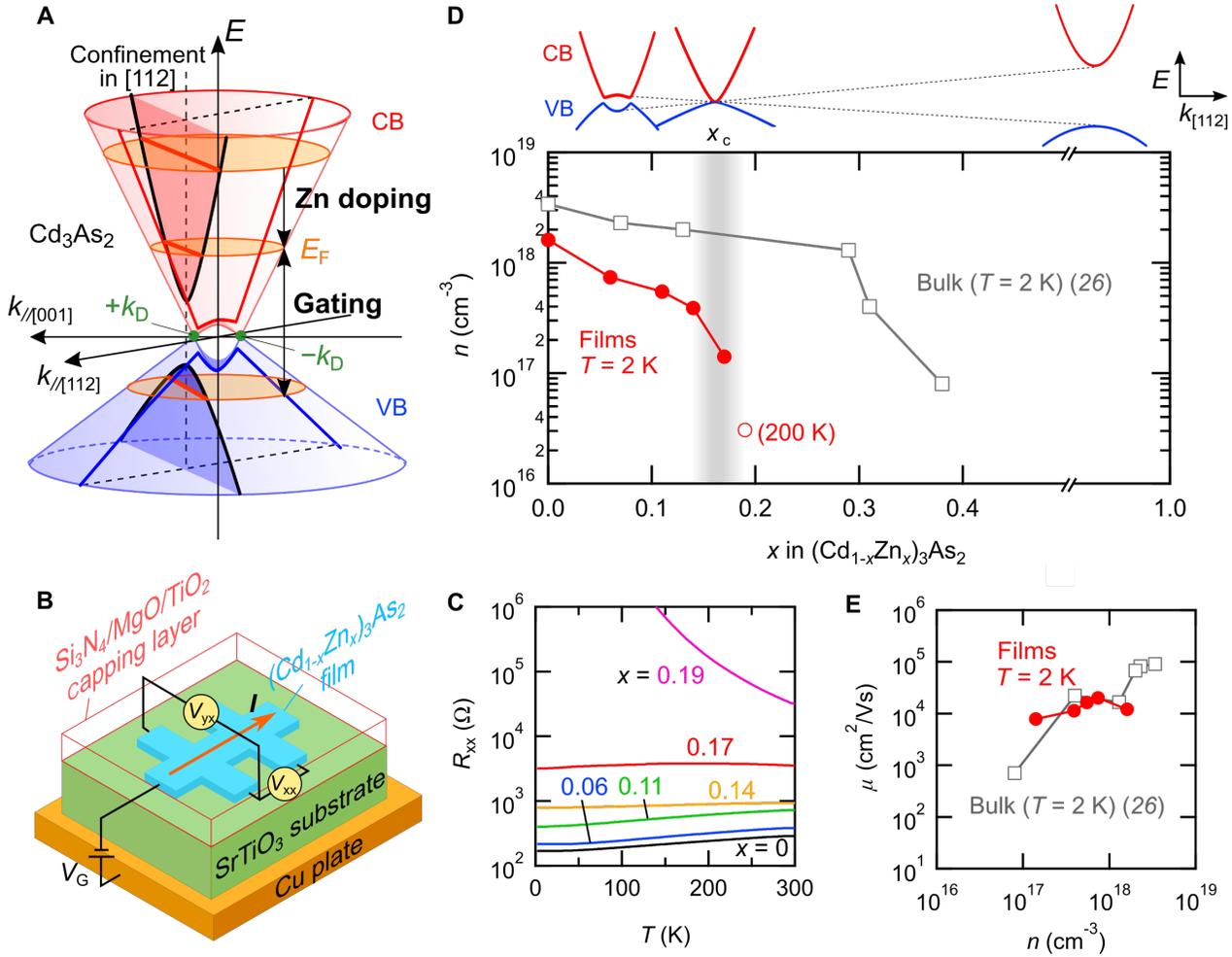

**Fig.1 Carrier control in $Cd_3As_2$ thin films by Zn doping and gating.** (**A**) Schematics of the band structure of two-dimensionally confined $Cd_3As_2$. The bulk Dirac points ($\pm k_D$) are on the [001] axis, while the growth direction of the thin films is along the [112] axis. Quantum confinement induces the formation of quantized sub-bands on the plane perpendicular to the [112] axis and the sub-band dispersion is described by a hyperbolic function. The carrier density is controlled by the combination of chemical Zn doping and electrostatic gating in a transistor configuration to investigate quantum transport at various Fermi level positions ($E_F$). (**B**) Schematic of back-gate transistor configuration using $SrTiO_3$ substrate as gate dielectric and a Cu plate as back-gate electrode. Carriers are depleted (accumulated) when the gate voltage $V_G$ is negatively (positively) biased. (**C**) Temperature dependence of resistance in thin films with different Zn concentration $x$. The film thicknesses are 35 nm. (**D**) Electron density vs. Zn concentration $x$ of thin films compared with of bulk single crystals. The thin films show a more efficient suppression of the electron density with smaller $x$ than in the case of bulk samples (*26*). The sketch on the top depicts the topological phase transition of $(Cd_{1-x}Zn_x)_3As_2$ from a Dirac semimetal to a trivial insulator. The critical doping level was estimated to be $x_c \sim 0.17$ (*28*). (**E**) Relationship of mobility and electron density of thin films and bulk samples (*26*).

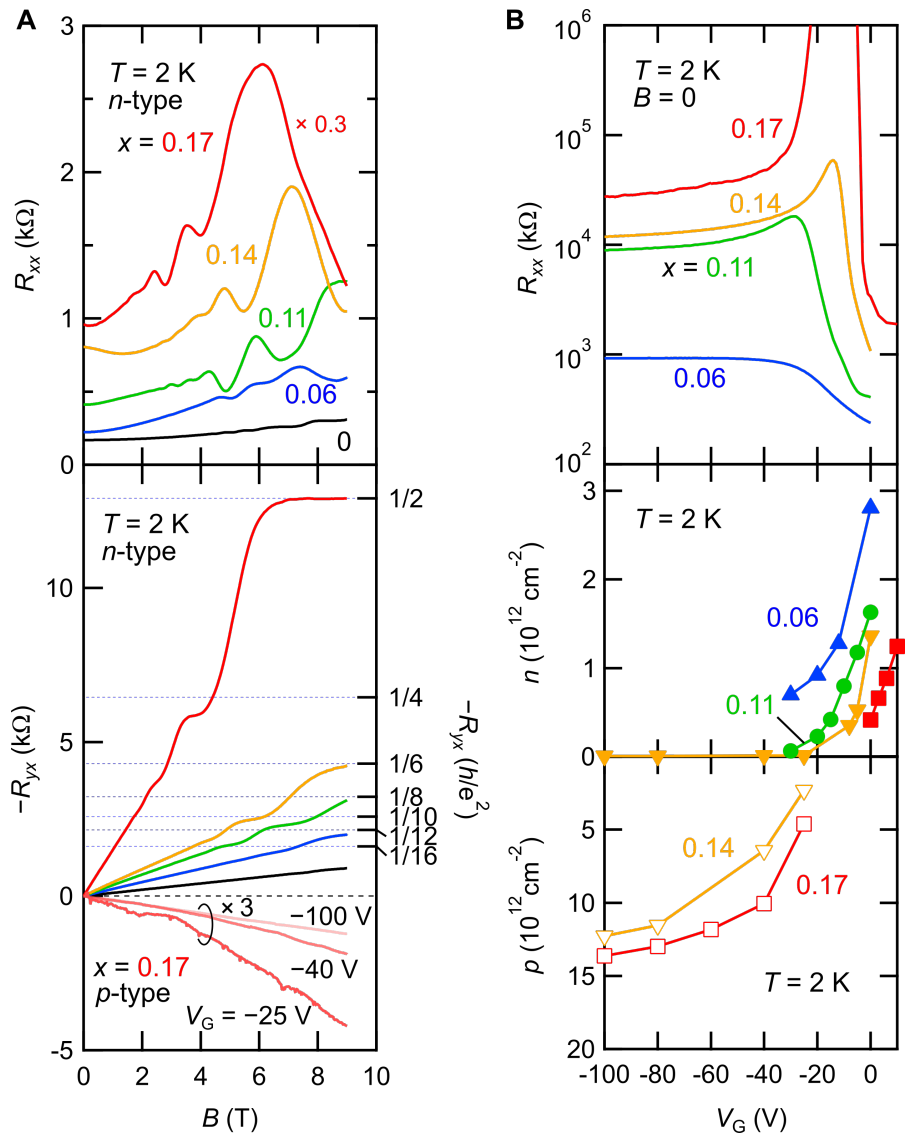

**Fig.2 Carrier-density control in $(Cd_{1-x}Zn_x)_3As_2$ thin films.** The film thicknesses are 35 nm. (**A**) Magnetic field dependence of longitudinal resistance $R_{xx}$ and Hall resistance $R_{yx}$ at 2 K. Quantized plateau regions appear in $R_{yx}$ as the electron density is decreased. Applying a large negative gate voltage $V_G$ induces a clear sign change of the Hall coefficient as shown for the case of $x = 0.17$. (**B**) $V_G$ dependence of $R_{xx}$, sheet electron density $n$, and sheet hole density $p$ at 2 K. Ambipolar behaviors are achieved by back-gating the samples in the case of $x \geq 0.11$. Accompanied with the carrier-type inversion, $R_{xx}$ exhibits a peak which corresponds to the charge neutrality point (CNP). In the cases of $x = 0.11$ and 0.14, multi-carrier conduction with majority holes and minority electrons is observed after reaching the CNP (see Fig. S2).

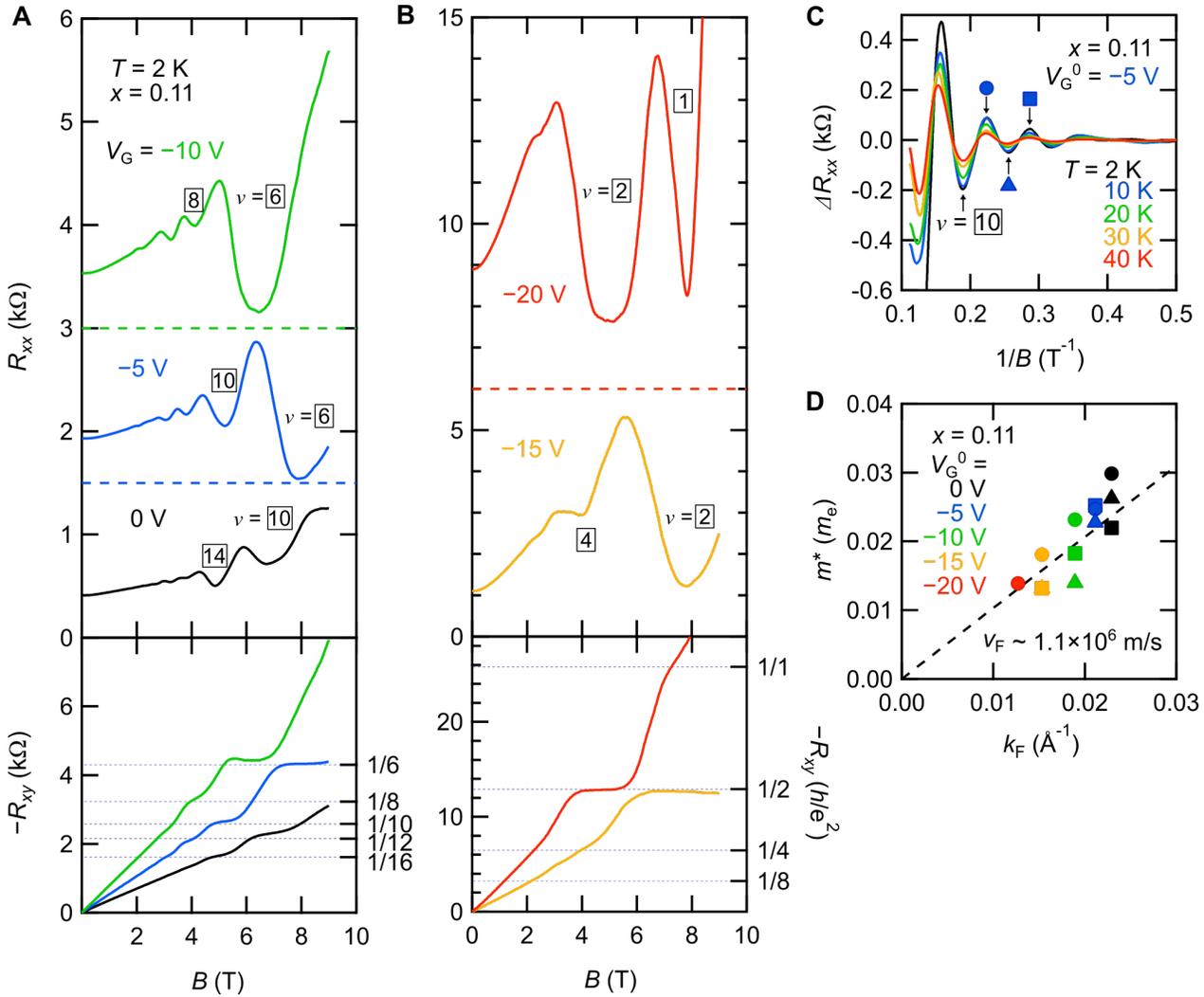

**Fig. 3**. **Gate-tunable quantum Hall states in $(Cd_{1-x}Zn_x)_3As_2$ film ($x = 0.11$).** (**A,B**) Magnetic field dependence of $R_{xx}$ and $R_{yx}$ measured at $V_G$ from 0 V to −20 V. The film thickness is 35 nm. QH states with smaller filling factors appear one by one as carriers are depleted. (**C**) Typical temperature-dependent oscillation terms of $R_{xx}$ at lower fields. $V_G^0$ denotes $V_G$ applied at 2 K and $V_G$ for higher temperatures is adjusted to keep the electron density constant. Here, the result for $V_G^0 = -5$ V is shown. (**D**) Relationship between the effective mass $m^*$ and the Fermi momentum $k_F$ of the main sub-band SB1. The different symbols represent $m^*$ values estimated using different oscillation peaks and valleys as shown in Fig. 3(C) for $V_G^0 = -5$ V (see Fig. S4(A) for the cases of other $V_G^0$ values). The Fermi velocity of SB1 is estimated to be about $1.1 \times 10^6$ m/s.

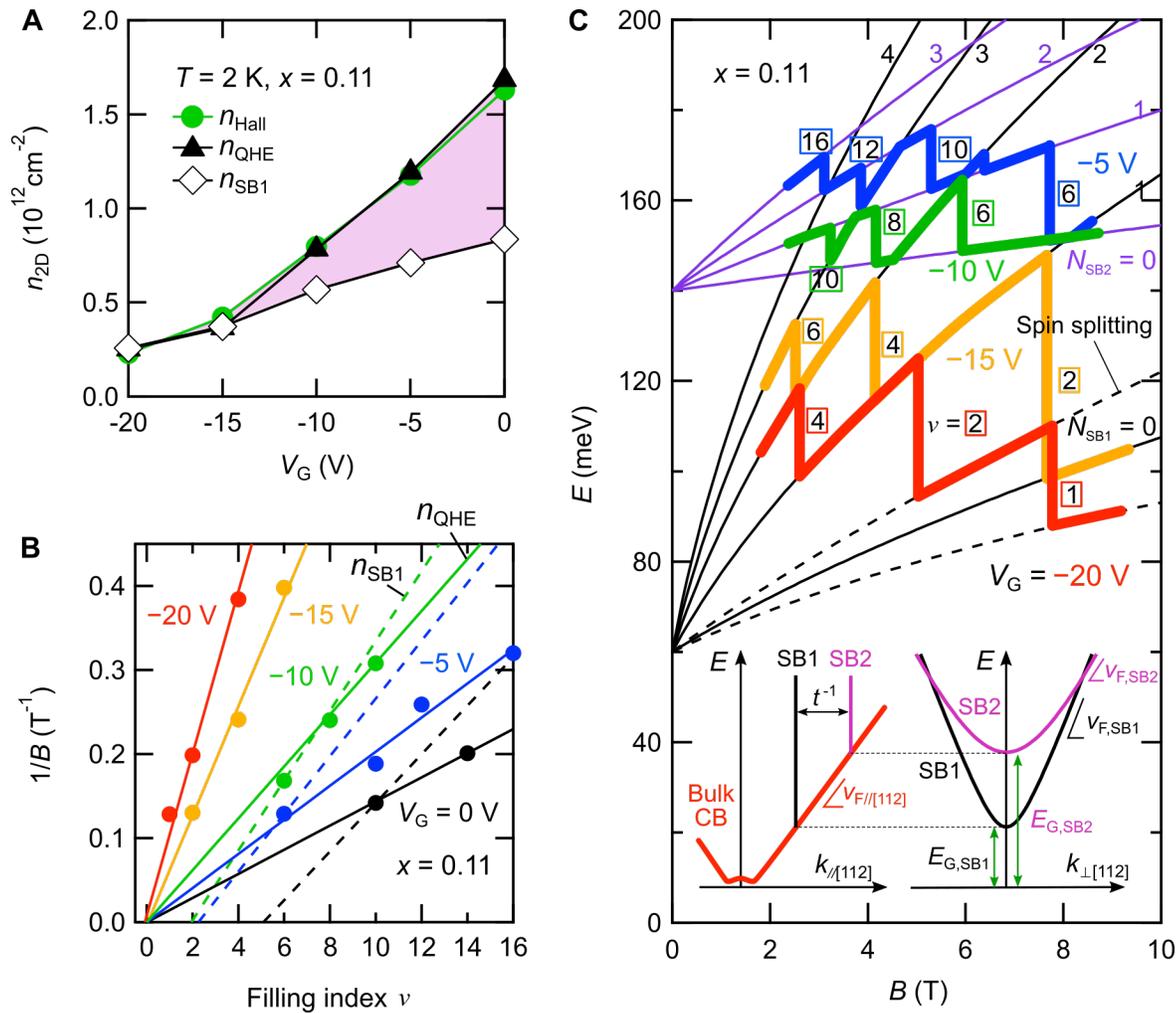

**Fig. 4 Landau level occupation identified from the quantum Hall states.** (**A**) Gate voltage $V_G$ dependence of carrier densities obtained from different experimental data. $n_{Hall}$ is obtained from Hall measurements, $n_{QHE}$ from the periodicity of the $R_{yx}$ quantization, and $n_{SB1}$ from the Fermi surface of the main sub-band SB1. While both $n_{Hall}$ and $n_{QHE}$ indicate the total electron density, $n_{SB1}$ corresponds to the partial electron density in SB1. The difference between the total electron density and $n_{SB1}$ at $V_G > -15$ V reflects the carrier population in the second sub-band SB2. (**B**) Filling factor plotted against the inverse of the magnetic field. The solid lines are fits to the data points with a fixed intercept at $\nu = 0$, yielding an estimate of $n_{QHE}$. The broken lines are drawn to pass each data point with Fermi surface area of SB1 as their slopes, representing the relationship between the filling factor and the magnetic field when only the presence of SB1 is assumed. The intercept of the broken lines can be used to estimate the Landau level occupation in SB2 as explained in the main text (see also Fig. S6). (**C**) Magnetic field dependence of Landau levels in SB1 and SB2, and their occupation transition at different Fermi levels. The thick lines overlaid on the Landau levels are the positions of chemical potential for each gate voltage. Accidental overlaps of the Landau levels in SB1 and SB2 result in jumps of the filling factor $\nu$ by 4. Inset is a sketch displaying the dispersions of the bulk conduction band along the confinement direction [112] (left), and those of sub-bands SB1, SB2 perpendicular to [112] (right). Each of the sub-band parameters is estimated to agree well with the observed plateau transitions. The sub-band splitting energy ($E_{G,SB2} - E_{G,SB1}$) reaches about 80 meV at a confinement thickness $t = 35$ nm, corresponding to the Fermi velocity $v_{F//[112]}$ of about $7\times10^5$ m/s.

**Supplementary Materials**

Supplementary materials are provided in a separate file.

S1. X-ray diffraction and estimation of Zn concentration $x$

Fig. S1. X-ray diffraction and estimation of Zn concentration $x$.

S2. Magnetotransport of $(Cd_{1-x}Zn_x)_3As_2$ films in $p$-type region

Fig. S2. Magnetotransport of $(Cd_{1-x}Zn_x)_3As_2$ films in $p$-type region.

S3. Temperature dependence of magnetotransport

Fig. S3. Temperature dependence of magnetotransport.

S4. Estimation of Fermi momentum and effective mass

Fig. S4. Estimation of Fermi momentum $k_F$.

Fig. S5. Estimation of effective mass $m^*$.

S5. Analysis of Landau level occupation

Fig. S6. Analysis of Landau level occupation.

Fig. S7. Appearance of quantum Hall states in the samples with different thicknesses.

S6. Estimation of sub-band parameters

Fig. S8. Determination of field positions for Hall plateau transitions.

Fig. S9. Estimation of sub-band parameters.